\documentclass{webofc}
\usepackage[varg]{txfonts}   
\usepackage{bm}
\def\Oc{{\rm O}} \def\S{{\rm S}}
\begin{document}
\title{Exotic bottomonium hadronic transitions}

\author{\firstname{Jaume} \lastname{Tarr\'us Castell\`a}\inst{1,3}\fnsep\thanks{\email{jtarrus@iu.edu}} }

\institute{Department of Physics, Indiana University, Bloomington, Indiana 47401, USA
\and  Center for Exploration of Energy and Matter, Indiana University, Bloomington, Indiana 47408, USA}

\abstract{We report on a recent computation of the transitions of exotic bottomonium to standard bottomonium and light quark hadrons. We work under the assumption that the $\Upsilon(10753)$ and $\Upsilon(11020)$ can be described as the lowest laying and first excitation $1^{--}$ hybrid bottomonium states, respectively. The computation has two distinct parts: the heavy quark transition matrix elements, which are obtained in a nonrelativistic EFT incorporating the heavy quark, multipole and adiabatic expansions; and the hadronization of the gluonic operators into the light-meson final states. The single mesons production is obtained through the axial anomaly and a standard $\pi^0-\eta-\eta'$ mixing scheme. Two pion and kaon production is obtained by solving the coupled Omnès problem. We also present result for semi-inclusive transitions.
}
\maketitle
\section{Introduction}\label{s1}

The nature of many of the exotic quarkonium states discovered so far it is still not settled. One of the difficulties in clarifying their nature is that many of the theoretical studies done so  far have been focused in the spectrum of these states. However, experimentally only a few $J^{PC}$ quantum numbers are easily accessible. This has lead, for instance, to a plethora of discoveries of $1^{--}$ exotic states, for which different models and approaches give similar predictions. Moreover, the composition of the heavy-quark spin symmetry multiplets, which could be used to distinguish different model pictures, cannot be tested due to this limitation in the accessible quantum numbers. On the other hand information on exotic quarkonium decay channels is considerably more abundant, since, at least, we know the decay channel in which the state has been discovered. Many of these decay channels are transitions into standard quarkonium and some light quark hadrons.  

\begin{figure}
\centerline{\includegraphics[width=0.5\textwidth]{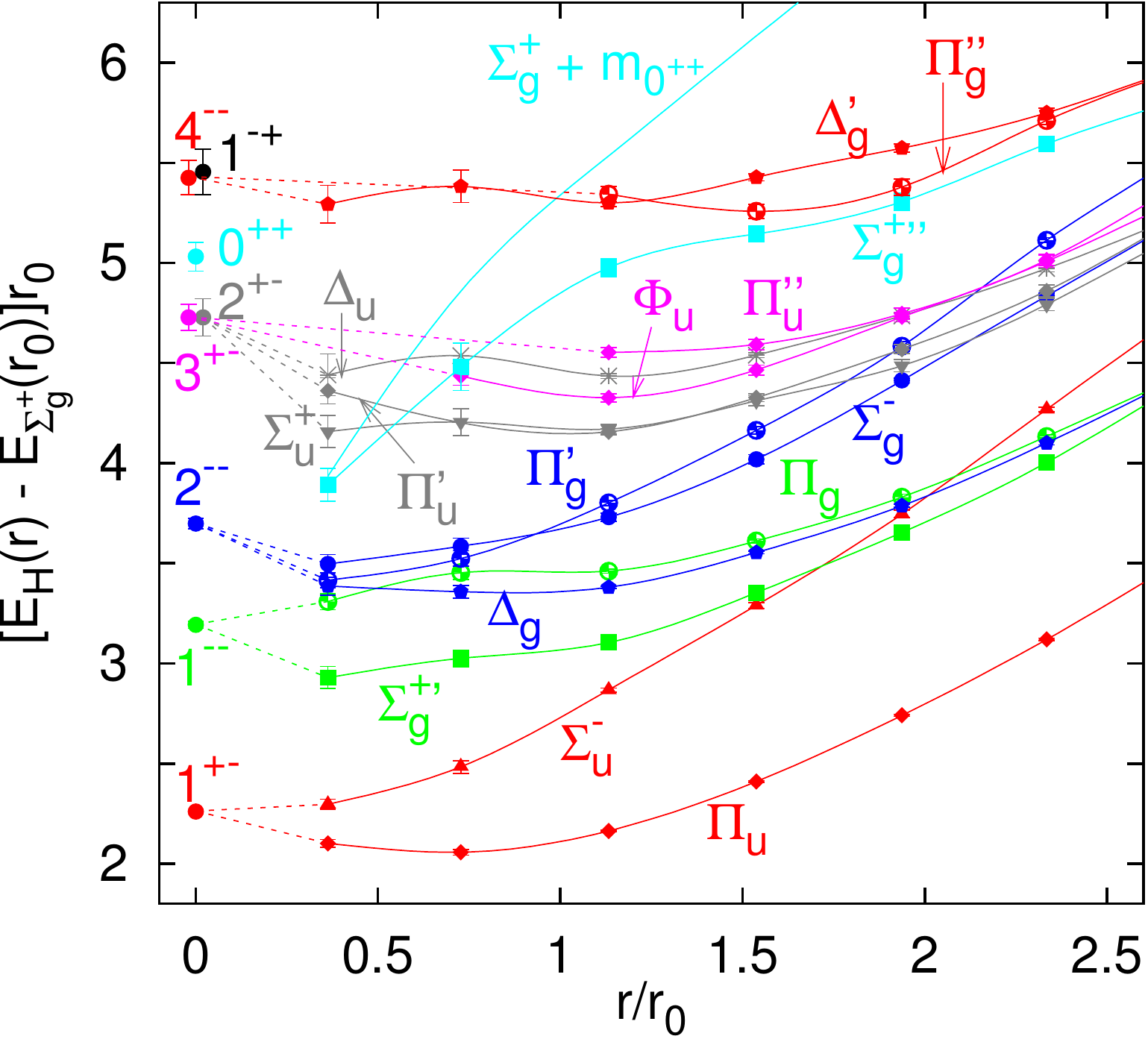}}
\caption{Quenched lattice NRQCD results for the heavy quark-antiquark static energy spectrum from Ref.~\cite{Juge:2002br}. Figure from Ref.~\cite{Bali:2003jq}.}
\label{fig-1}
\end{figure}

An effective field theory (EFT) description of exotic quarkonium has been developed in past few years. This EFT is build upon two expansions. The first one is the heavy-quark mass expansion. Therefore, a natural starting point is NRQCD at leading order, that is in the static limit. The energy spectrum of a heavy quark-antiquark pair in the static limit is formed by the static energies (also referred to as adiabatic surfaces). These are the energies of the eigenstates of the leading order NRQCD Hamiltonian for a quark-antiquark system. These eigenstates are characterized by a set of quantum numbers~\cite{Soto:2020xpm}: the total spin, parity and charge conjugation of the light degrees of freedom, i.e. the glue and light-quark content of the exotic quarkonium state; the light-quark flavor; and quantum numbers labeling the representation of $D_{\infty h}$ (see for instance Ref.~\cite{Berwein:2015vca}). The latter being the cylindrical symmetry group describing the spatial symmetries of the static heavy quark-antiquark system. The static energies must be computed using nonperturbative techniques. The most well known case is for heavy quark-antiquark pairs with isospin $I=0$ light degrees of freedom where several computations of the static energy spectrum in the quenched approximation are available in Refs.~\cite{Bali:2000vr,Juge:2002br,Capitani:2018rox,Schlosser:2021wnr}. The results of Ref.~\cite{Juge:2002br} are displayed in Fig.~\ref{fig-1}. The states supported by such static energies correspond to the quarkonium hybrid picture.

The second expansion incorporated in our EFT framework is the adiabatic expansion between the heavy-quark dynamics and that of the light degrees of freedom. When one goes beyond the static limit in the heavy quark mass expansion, one can find heavy quark-antiquark bound states supported by the static energies in a picture analogous to that of the Born-Oppenheimer approximation for diatomic molecules~\cite{Brambilla:2017uyf}. In fact the EFT at leading order reproduces the naive use of Born-Oppenheimer approximation  for exotic quarkonium systems~\cite{Griffiths:1983ah,Juge:1999ie}. Due to this, the EFT is often refereed as Born-Oppenheimer EFT (BOEFT). For hybrid quarkonium systems BOEFT has been developed up to $1/m^2_Q$ heavy quark spin dependent terms in \cite{Berwein:2015vca,Brambilla:2017uyf,Oncala:2017hop,Brambilla:2018pyn,Brambilla:2019jfi}. A general formulation for any light quark content has been obtained more recently~\cite{Soto:2020xpm} which can also be used for heavy quark-quark systems such as doubly heavy baryons~\cite{Soto:2020pfa,Soto:2021cgk}. Combining the hybrid quarkonium spectrum with the standard quarkonium states close and above open flavor thresholds one can account for the observed exotic quarkonium states~\cite{Oncala:2017hop}. A dominant molecular component for some of these states can be explained through their coupling to heavy meson-antimeson pairs~\cite{TarrusCastella:2022rxb}. Exotic states corresponding to four heavy quark resonances are not expected to be described by the BOEFT framework, nevertheless other EFT approaches are available for such case~\cite{Brambilla:2017ffe}.

\begin{figure}
\centerline{\includegraphics[width=0.8\textwidth]{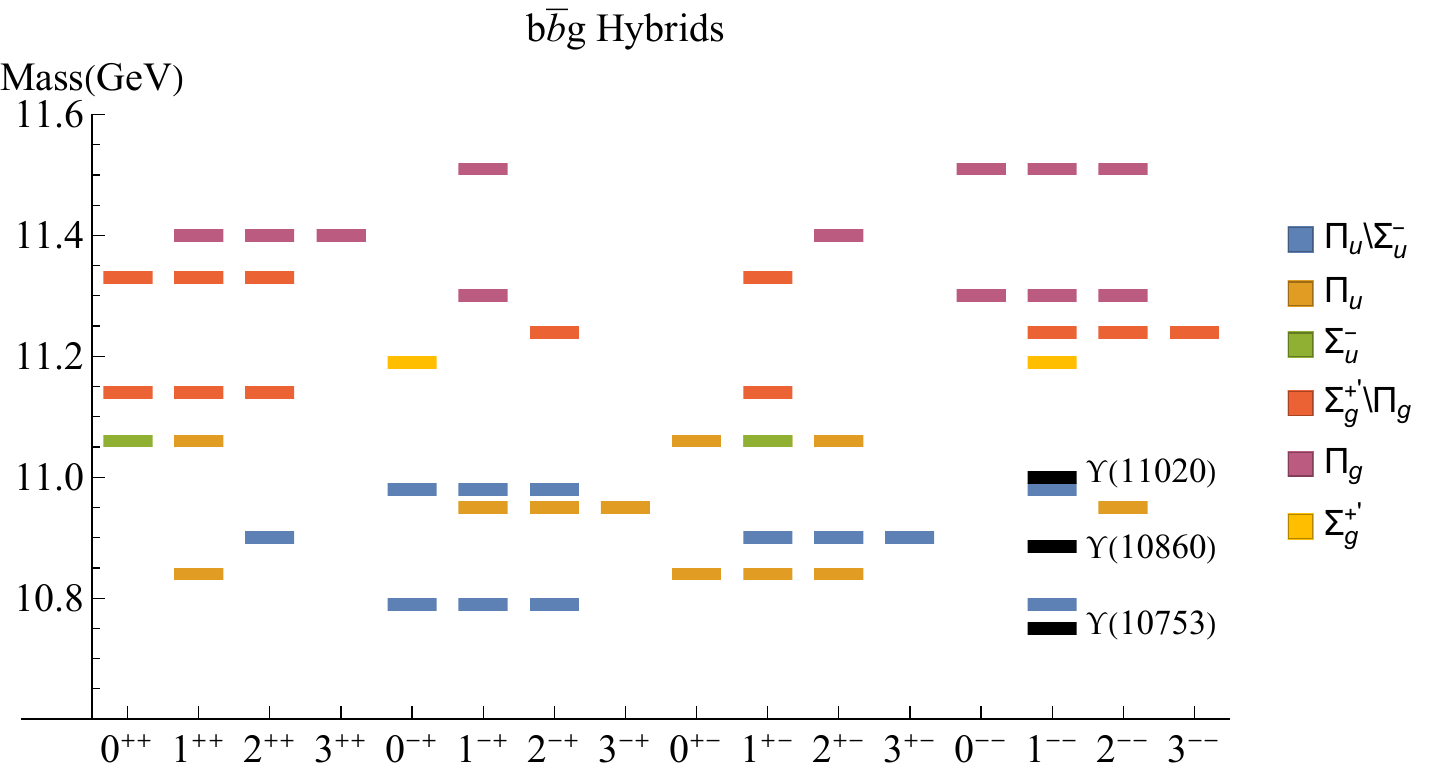}}
\caption{Bottomonium hybrid spectrum from Refs.~\cite{Berwein:2015vca,Pineda:2019mhw}. Each line represent a state. The color of each line indicates the static energies that contribute to each state. The black lines correspond to the three neutral exotic bottomonium states discovered so far.}
\label{fig-2}
\end{figure}

The spectrum of bottomonium hybrids at leading order in BOEFT supported by the lowest laying static energy multiplet ($\Pi_u-\Sigma_u^-$)~\cite{Berwein:2015vca} and the next to lowest laying multiplet ($\Sigma_g^{+\prime}-\Pi_g$)~\cite{Pineda:2019mhw} is shown in Fig.~\ref{fig-2}. Additionally in the figure we show the three known neutral exotic bottomonium states as black lines. It is interesting that the $\Upsilon(10753)$ and $\Upsilon(11020)$ fit quite nicely into the predictions for ground and first excited $1^{--}$ bottomonium hybrid states. In particular the experimental masses are compatible with the theoretical predictions considering the uncertainty. This result motivated us to study the transitions of $\Upsilon(10753)$ and $\Upsilon(11020)$ under the assumption that these states are bottomonium hybrids, work  which we presented in Ref.~\cite{TarrusCastella:2021pld}. To study the transitions of hybrid to standard bottomonium states we use weakly coupled pNRQCD (pNRQCD)~\cite{Pineda:1997bj,Brambilla:1999xf} in a similar approach to the one used in Ref.~\cite{Pineda:2019mhw} to study transitions in standard quarkonium. This is equivalent to the use of the multipole expansion in the writing of the transition operators. Since the multipole expansion is only valid for $r\lesssim 1/\Lambda_{\rm QCD}$ this implies that we are working in the short heavy quark-antiquark distance regime. Due to this we do not extend our study to the charmonium sector.

\section{Hybrid and standard quarkonium states} \label{s2}

The standard quarkonium states in the static limit read as
\begin{align}
|{\bf R}, {\bf r};\Sigma_g^+\rangle =\S^{\dagger}\left(\bm{R},\bm{r}\right)|0\rangle\,,
\end{align}
with $\S$ the heavy quark-antiquark singlet field~\cite{Brambilla:1999xf}. The corresponding full static potential matches the $\Sigma_g^+$  static energy 
\begin{align}
V^{(0)}_{\Sigma_g^+}(r)&=\lim_{t \rightarrow \infty}\frac{i}{t}\ln \langle{\bf R}, {\bf r};\Sigma_g^+; t/2 |{\bf R}, {\bf r};\Sigma_g^+;-t/2\rangle=V^{(0)}_s+b_{\Sigma_g^+} r^2+\cdots=E^{(0)}_{\Sigma_g^+}(r)\,.
\end{align}
A basis for a general quarkonium state can be build from the static states and $\phi^{(m)}(\bm{r})$, the quarkonium wave function,
\begin{align}
&|S_m\rangle=\int d^3\bm{r}d^3\bm{R}\,\phi^{(m)}(\bm{R}, \bm{r})|\bm{R}, \bm{r};\Sigma_g^+\rangle\,.
\end{align}
Using quantum mechanical perturbation theory we can incorporate the kinetic term and obtain the Shr\"odinger equation for the standard quarkonium states
\begin{align}
&\left(-\frac{\bm{\nabla}^2_r}{m_Q}+V^{(0)}_{\Sigma_g^+}(r)\right)\phi^{(m)}(\bm{r})={\cal E}_m \phi^{(m)}(\bm{r})\,.
\end{align}

The construction of the hybrid quarkonium states is slightly more complicated due to their nontrivial gluonic content. The lowest laying hybrid states correspond to the $\Sigma_u^-$ and $\Pi_u$ static energies, which in the short distance can be constructed as the projections into the heavy quark-antiquark axis of a glue operator, called gluelump, with quantum numbers $1^{+-}$. The gluelump operator can be expanded into all the glue operators with matching quantum numbers. Then we assume that there is a correlation between the dimensionality of the interpolating operator and the strength of the interpolation with the hybrid, such that higher dimension operators are subleading~\cite{Pineda:2019mhw}, so the series can be truncated at LO. This hypothesis is supported by the ordering of the hybrid static energies. Therefore, the gluelump operator can be approximated as $\bm{G}_{B}^{a}\sim  Z^{-1/2}_B\bm{B}^a$. One can estimate the value of $Z_B$ using the normalization of the gluelump operators to relate it to the value of the gluon condensate. The latter is taken from Ref.~\cite{Ayala:2020pxq}.

The Hybrid static states containing a $1^{+-}$ gluelump can be written as
\begin{align}
|{\bf R}, {\bf r};\lambda\rangle =\bm{\hat{r}}_{\lambda}\cdot\bm{G}_{B}^{a}(\bm{R})O^{a\,\dagger}\left(\bm{R},\bm{r}\right)|0\rangle,
\end{align}
with $O^a$ the heavy quark-antiquark octet field. The full static potentials correspond to
\begin{align}
V^{(0)}_{\lambda}(r)&=\lim_{t \rightarrow \infty}\frac{i}{t}\log \langle{\bf R}, {\bf r};\lambda; t/2 |{\bf R}, {\bf r};\lambda;-t/2\rangle=E^{(0)}_{|\lambda|}(r)\,,
\end{align}
with $E^{(0)}_{0}(r)=E^{(0)}_{\Sigma_u^-}(r)$ and $E^{(0)}_{|\pm 1|}(r)=E^{(0)}_{\Pi_u}(r)$ obtained from lattice QCD. To go beyond the static limit we use that an eigenstate of the full Hamiltonian can be expressed in the basis of eigenstates of the static limit
\begin{align}
|H_n\rangle=\int d^3\bm{r}d^3\bm{R}\sum_{\lambda}\psi^{(n)}_{\lambda}(\bm{R}, \bm{r})|\bm{R}, \bm{r};\lambda\rangle\,.
\end{align}
Using quantum mechanical perturbation theory to incorporate the kinetic operator one arrives at the coupled Shr\"odinger equations for the hybrid bound states
\begin{align}
\sum_{\lambda}\left(-\bm{\hat{r}}^{*}_{\lambda'}\frac{\bm{\nabla}^2_r}{m_Q}\bm{\hat{r}}_{\lambda}+V^{(0)}_{\lambda}(r)\delta_{\lambda'\lambda}\right)\psi^{(n)}_{\lambda}(\bm{r})={\cal E}_n \psi^{(n)}_{\lambda'}(\bm{r})\,.
\end{align}

\section{Exclusive Transitions} \label{s3}

In weakly-coupled pNRQCD the transitions between  hybrid and standard quarkonium are generated by the operators that couple the heavy quark pair singlet and octet fields. These operators start appearing at NLO in the multipole or heavy quark mass expansions. The two lowest order transition operators read as follows:
\begin{align}
L_{\rm pNRQCD} =\int d^3R d^3r\,\Bigg\{&g{\rm Tr}\left[\S^{\dag}\bm{r}\cdot\bm{E}\,\Oc+\Oc^{\dag}\bm{r}\cdot\bm{E}\,\S\right]\nonumber \\
&+\frac{gc_F}{m_Q}{\rm Tr}\left[\S^{\dag}(\bm{S}_1-\bm{S}_2)\cdot\bm{B}\,\Oc+\Oc^{\dag}(\bm{S}_1-\bm{S}_2)\cdot\bm{B}\,\S\right]\Bigg\}\,.
\label{pnrqcd2}
\end{align}  

The spin vectors $\bm{S}_1$ and $\bm{S}_2$ correspond to the heavy-quark and heavy-antiquark respectively. The chromoelectric and chromomagnetic fields are defined as $\bm{E}^i=G^{i0}$ and $\bm{B}^i=-\epsilon_{ijk}G^{jk}/2$ with $\epsilon_{123}=1$.

Let us compute the transition generated by the first (leading order) operator in Eq.~\eqref{pnrqcd2}. The amplitude is as follows
\begin{align}
\langle S_{m}{\cal O}_{\pi}|g{\rm Tr}\left[\S^{\dag}\bm{r}\cdot\bm{E}\,\Oc\right]|H_n\rangle=\frac{1}{3}\sqrt{\frac{T_F}{N_cZ_B}}\langle {\cal O}_{\pi}|g^2\bm{E}\cdot\bm{B}| 0\rangle \int d^3r\sum_{\lambda}\phi^{(m)}(\bm{r})\bm{r}\cdot\bm{\hat{r}}_\lambda\psi^{(n)}_{\lambda}(\bm{r})\,,\label{amp1}
\end{align} 
 with ${\cal O}_{\pi}$ denoting a generic final light-quark state. As can be seen from Eq.~\eqref{amp1}, the amplitude factorizes into some constant factors, a heavy quark matrix element and a  gluonic matrix element. Selection rules can be derived from the wave functions integral in the heavy quark matrix element. Since the transition operator is independent of the heavy-quark spin this should be conserved. If we identify $\Upsilon(10753)$ and $\Upsilon(11020)$ as hybrid bottomonium with $n^1{\cal P}_1$ and $n=1,2$, respectively, then the final quarkonium states must be $h_b(m^1P_1)$. The gluonic operator has quantum numbers $0^{-+}$ and isospin $I=0$, therefore the allowed final light-quark states must match these quantum numbers. Some examples of these states are $\pi^0$, $\eta$, $\eta'$, higher mass $\eta$-like resonances or odd numbers of mesons such as $\pi^0\pi^+\pi^-$ or $\eta\,\pi^+\pi^-$. 

We compute the gluonic matrix element for the production of $\pi^0$, $\eta$, $\eta'$. These matrix elements can be determined from $U(1)_A$ anomaly and a mixing scheme. The gluonic matrix element in Eq.~\eqref{amp1} can be rewritten as $(g^2/\pi \bm{E}\cdot\bm{B})=\alpha_s G_{\mu\nu}\tilde{G}^{\mu\nu}$ with the dual field-strength tensor defined as $\tilde{G}^{\mu\nu}=\frac{1}{2}\epsilon^{\mu\nu\alpha\beta}G_{\alpha\beta}$ and $\epsilon_{0123}=1$. The matrix element to obtain is then
\begin{align}
\omega_c=\langle 0|\frac{\alpha_s}{4\pi}G_{\mu\nu}\tilde{G}^{\mu\nu}|\eta_{c}(p)\rangle\,,\quad c=\pi^0,\eta,\eta'\label{Ap:krollm:e2}\,.
\end{align}
The matrix elements of $G_{\mu\nu}\tilde{G}^{\mu\nu}$ can then be related to the divergence of the axial current and the pseudoscalar current through the axial anomaly. This leaves us with $18$ nonperturbative parameters corresponding to the matrix elements of the axial and pseudoscalar currents and final states $\pi^0, \eta, \eta'$. This amount of free parameters can be greatly reduced by the implementation of a mixing scheme between $\pi^0-\eta-\eta'$. We use the one in Refs.~\cite{Feldmann:1998vh,Kroll:2005sd}. The remaining parameters can be obtained from the masses and decay widths of the Goldstone bosons.

We obtain the following decay widths for the transitions of $\Upsilon(10753)$ and $\Upsilon(11020)$ to $h_b(nP)$ and $\pi^0,\,\eta$ or $\eta'$ in the final state:
\begin{align}
&\Gamma_{\Upsilon(10753)\to h_b(1P)\pi^0}=2.57(\pm 1.03)_{\rm m.e.}(\pm 0.14)_{Z_B}(\pm 0.16)_{\omega_{\pi^0}}~{\rm keV}\,,\label{s2:t1}\\
&\Gamma_{\Upsilon(10753)\to h_b(1P)\eta}=2.29(\pm 0.92)_{\rm m.e.}(\pm 0.13)_{Z_B}(\pm 0.08)_{\omega_{\eta}}~{\rm MeV}\,,\label{s2:t2}\\
&\Gamma_{\Upsilon(10753)\to h_b(2P)\pi^0}=0.168(\pm 0.067)_{\rm m.e.}(\pm 0.009)_{Z_B}(\pm 0.010)_{\omega_{\pi^0}}~{\rm keV}\,,\label{s2:t3}\\
&\Gamma_{\Upsilon(11020)\to h_b(1P)\pi^0}=2.04(\pm 0.82)_{\rm m.e.}(\pm 0.11)_{Z_B}(\pm 0.13)_{\omega_{\pi^0}}~{\rm keV}\,,\label{s2:t4}\\
&\Gamma_{\Upsilon(11020)\to h_b(1P)\eta}=2.04(\pm 0.81)_{\rm m.e.}(\pm 0.11)_{Z_B}(\pm 0.07)_{\omega_{\eta}}~{\rm MeV}\,,\label{s2:t5}\\
&\Gamma_{\Upsilon(11020)\to h_b(1P)\eta'}=9.23(\pm 3.69)_{\rm m.e.}(\pm 0.51)_{Z_B}(\pm 0.39)_{\omega_{\eta'}}~{\rm MeV}\,,\label{s2:t6}\\
&\Gamma_{\Upsilon(11020)\to h_b(2P)\pi^0}=0.104(\pm 0.042)_{\rm m.e.}(\pm 0.006)_{Z_B}(\pm 0.006)_{\omega_{\pi^0}}~{\rm keV}\,,\label{s2:t7}\\
&\Gamma_{\Upsilon(11020)\to h_b(2P)\eta}=81.8(\pm 32.7)_{\rm m.e.}(\pm 4.6)_{Z_B}(\pm 2.7)_{\omega_{\eta}}~{\rm keV}\,.\label{s2:t8}
\end{align}
The uncertainties are labeled by their origin. The largest source of uncertainty is the use of the multipole expansion (${\rm m.e.}$). We estimate this uncertainty as corrections of ${\cal O}\left(\Lambda^2_{\rm QCD}r^2\right)$. 

Next we focus our attention into the transitions generated by the heavy quark mass suppressed operator in the Lagrangian in Eq.~\eqref{pnrqcd2}. Computing the expected value of this operator between an initial hybrid bottomonium state and a final standard bottomonium state plus some light quark hadrons (generically denoted by ${\cal O}_{\pi\pi}$) we obtain the following amplitude
\begin{align}
&\langle S_{m}{\cal O}_{\pi\pi}|\frac{gc_F}{m_Q}{\rm Tr}\left[\S^{\dag}(\bm{S}_1-\bm{S}_2)\cdot\bm{B}\,\Oc\right]|H_n\rangle\nonumber \\
&=\frac{gc_F}{3m_Q}\sqrt{\frac{T_F}{N_cZ_B}}\langle {\cal O}_{\pi\pi}|\bm{B}^2| 0\rangle \int d^3r\sum_{\lambda}\phi^{(m)}(\bm{r})(\bm{S}_1-\bm{S}_2)\cdot\bm{\hat{r}}_\lambda\psi^{(n)}_{\lambda}(\bm{r})\,,\label{tt2s1}
\end{align}  
As in the leading order transition the amplitude factorizes into a set of constant factors, a heavy quark matrix element and a light-quark hadron production matrix element. The heavy quark matrix element can be computed from the wave functions of the hybrid and standard bottomonium states involved in the transition. From the heavy-quark spin structure of the operator we can find the selection rules $\delta S=1$. Since in the hybrid bottomonium picture we are employing for the $\Upsilon(10753)$ and $\Upsilon(11020)$, these correspond to spin singlet hybrids, the final quarkonium must be spin triplet. Moreover the total $J^{PC}$ must be conserved. Therefore the final quarkonium states can only be $\Upsilon(m^3S_1)$ or $\Upsilon(m^3D_1)$. We will only consider the first case since $D$-wave bottomonium states have not yet been observed experimentally. 

\begin{figure}
\centerline{\includegraphics[width=.55\textwidth]{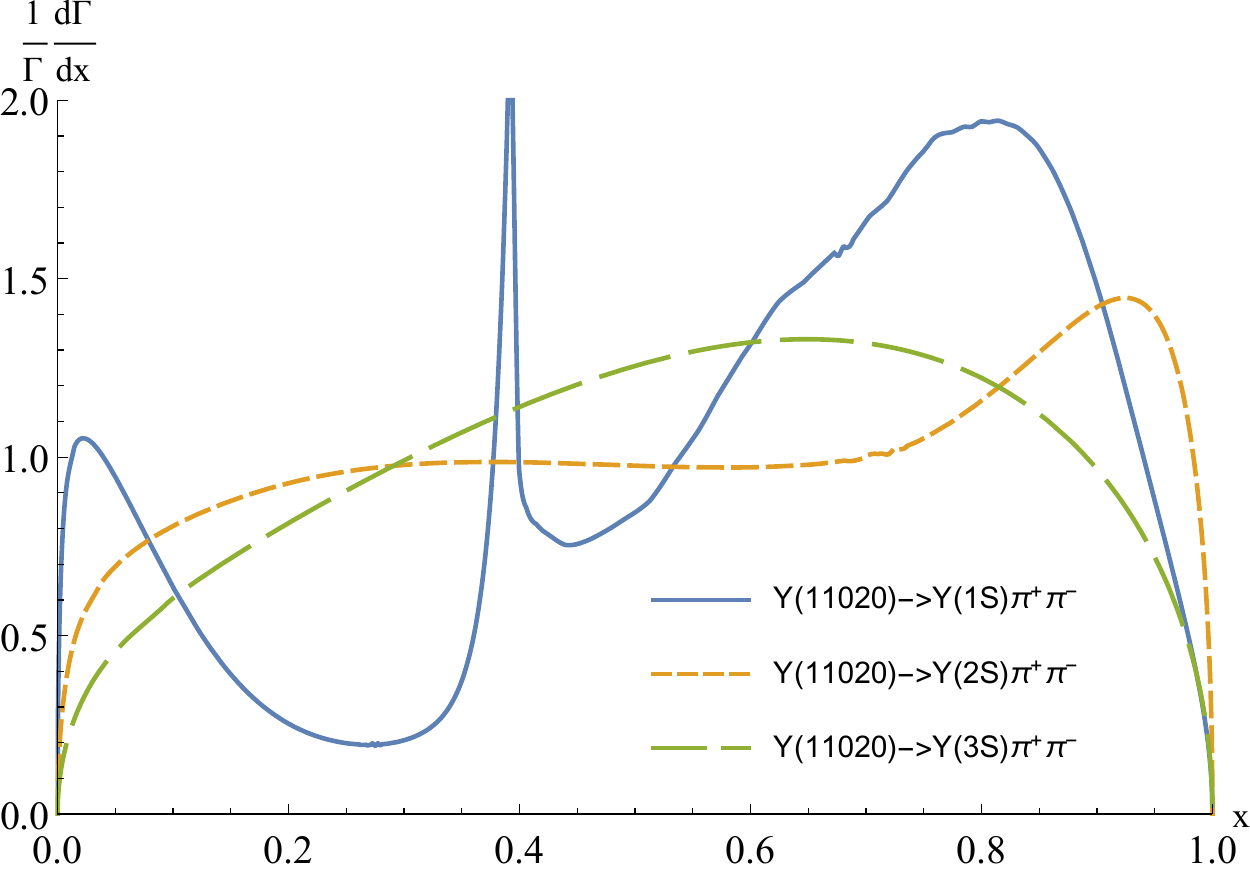} }
\caption{Normalized differential width for the transitions $\Upsilon(11020)\to \Upsilon(m^3S_1)\pi^+\pi^-$. The variable $x$ is defined as $x=(s-4m^2_{\pi})/(m_{\Upsilon(11020)}-m_{\Upsilon(mS)}-4m^2_{\pi})$.}
\label{lshpp}
\end{figure}

The allowed light-quark hadron products of the transition are controlled by the matrix element of the gluonic operator $\bm{B}^2$. Therefore the light-quark hadron products must be $0^{++}$ and isospin $I=0$. Such states are, for instance $\pi^+\pi^-$, $K^+K^-$, pairs of $\pi^0$ or $\eta$ as well as $f_0$ resonances up to the invariant mass allowed by the specific initial and final heavy-quark states. We compute the first two cases, $\pi^+\pi^-$, $K^+K^-$, for which we build a dispersive representation of the gluonic matrix element following Refs.~\cite{Donoghue:1990xh,Moussallam:1999aq}. However, unlike those references our matrix element contains not only an $S$-wave piece but also $D$-wave one. We have extended the coupled Muskhelishvili-Omn\`es approach to the $D$-wave final state interactions for the first time. We use the parametrizations of the $\pi\pi\to \pi\pi$ and $\pi\pi\to K\bar{K}$ partial waves from Refs.~\cite{GarciaMartin:2011cn,Pelaez:2018qny}, which to our knowledge are the most accurate currently available. For the numerical solution of the coupled Muskhelishvili-Omn\`es equations we use the techniques of Refs.~\cite{Moussallam:1999aq,Descotes-Genon:2000pfd}. The subtraction constants of the dispersive representations are determined by matching to a chiral representation of the form factors for small invariant mass of the dipion ($s$). This depends of three low-energy constants that can be determined from the scale anomaly~\cite{Chivukula:1989ds}, the Feynman-Helmann theorem and the last remaining one can be extracted from two pion transitions in standard quarkonium~\cite{Pineda:2019mhw}. In Fig.~\ref{lshpp} we plot the normalized differential decay widths for the transitions of $\Upsilon(11020)$ with to standard bottomonium and $\pi^+\pi^-$. Integrating the differential transition width over the kinematically allowed range of $s$ we obtain the following transition widths:
\begin{align}
\Gamma_{\Upsilon(10753)\to \Upsilon(1S)\pi^+\pi^-}&=43.4(\pm 17.3)_{\rm m.e.}(\pm 2.4)_{Z_B}(\pm 8.6)_{\alpha_s}(^{+0.5}_{-0.0})_{\kappa}~{\rm keV}\,,\label{s3:e1}\\
\Gamma_{\Upsilon(10753)\to \Upsilon(2S)\pi^+\pi^-}&=2.75(\pm 1.10)_{\rm m.e.}(\pm 0.15)_{Z_B}(\pm 0.55)_{\alpha_s}(^{+0.13}_{-0.12})_{\kappa}~{\rm keV}\,,\label{s3:e2}\\
\Gamma_{\Upsilon(10753)\to \Upsilon(3S)\pi^+\pi^-}&=0.98(\pm 0.39)_{\rm m.e.}(\pm 0.05)_{Z_B}(\pm 0.19)_{\alpha_s}(\pm 0.03)_{\kappa}~{\rm eV}\,,\label{s3:e3}\\
\Gamma_{\Upsilon(10753)\to \Upsilon(1S)K^+K^-}&=3.98(\pm 1.59)_{\rm m.e.}(\pm 0.22)_{Z_B}(\pm 0.79)_{\alpha_s}(^{-0.50}_{+0.67})_{\kappa}~{\rm keV}\,,\label{s3:e7}
\end{align}

and 
\begin{align}
\Gamma_{\Upsilon(11020)\to \Upsilon(1S)\pi^+\pi^-}&=99.1(\pm 39.6)_{\rm m.e.}(\pm 5.5)_{Z_B}(\pm 19.7)_{\alpha_s}(^{+26.3}_{-21.8})_{\kappa}~{\rm keV}\,,\label{s3:e4}\\
\Gamma_{\Upsilon(11020)\to \Upsilon(2S)\pi^+\pi^-}&=3.96(\pm 1.58)_{\rm m.e.}(\pm 0.22)_{Z_B}(\pm 0.70)_{\alpha_s}(^{-0.16}_{+0.17})_{\kappa}~{\rm keV}\,,\label{s3:e5}\\
\Gamma_{\Upsilon(11020)\to \Upsilon(3S)\pi^+\pi^-}&=1.33(\pm 0.53)_{\rm m.e.}(\pm 0.07)_{Z_B}(\pm 0.27)_{\alpha_s}(\pm 0.02)_{\kappa}~{\rm keV}\,,\label{s3:e6}\\
\Gamma_{\Upsilon(11020)\to \Upsilon(1S)K^+K^-}&=5.93(\pm 2.37)_{\rm m.e.}(\pm 0.33)_{Z_B}(\pm 1.18)_{\alpha_s}(^{+1.75}_{-1.18})_{\kappa}~{\rm keV}\,.\label{s3:e8}
\end{align}

As in the leading order transitions the uncertainties are labeled by their origin. The last two uncertainties are related to the determination of the low-energy constants of the chiral representation of the form factors.

\section{Semi-Inclusive Transitions}

\begin{figure}
\centerline{\includegraphics[width=.6\textwidth]{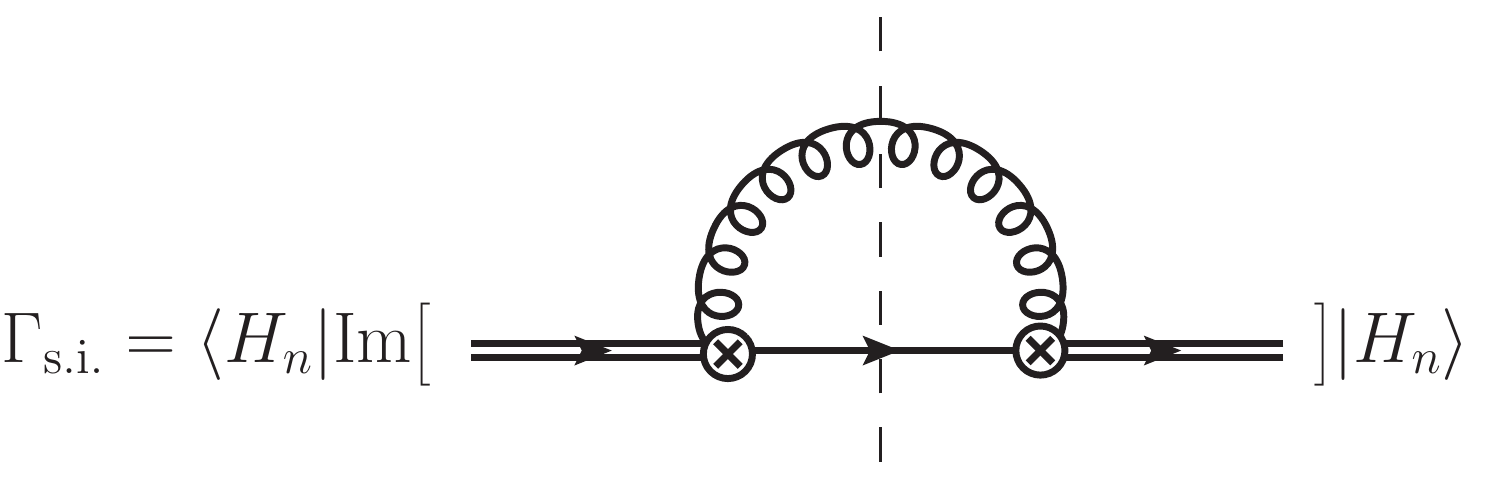}}
\caption{The single and double lines represent quarkonia in singlet and octet states respectively. The curly line stands for a gluon. Note that the spectator gluons forming the $H_n$ state are not displayed.}
\label{tptf}
\end{figure}

When the energy gap between a hybrid and a standard quarkonium state is large, the gluon emitted by the heavy quarks in the transition from an octet to a singlet state can be considered perturbative and semi-inclusive decay widths can be computed~\cite{Oncala:2017hop}. These semi-inclusive decay widths correspond to the expected value of the hybrid states of the imaginary part of the diagram in Fig.~\ref{tptf}. The vertices in the diagram can be either of the operators in the Lagrangian in Eq.~\eqref{pnrqcd2}. Among the transitions we studied in the exclusive channels, the following have large enough energy gaps
\begin{align}
&\Gamma^{\rm LO}_{\Upsilon(11020)\to h_b(1P)}=20(\pm 9)_{\alpha_s}~{\rm MeV}\,,\\
&\Gamma^{\rm NLO}_{\Upsilon(10753)\to \Upsilon (1S)}=9.7(\pm 3.8)_{\alpha_s}~{\rm MeV}\,,\label{s:siw:e2}\\
&\Gamma^{\rm NLO}_{\Upsilon(11020)\to \Upsilon (1S)}=7.3(\pm 2.5)_{\alpha_s}~{\rm MeV}\,,\label{s:siw:e3}\\
&\Gamma^{\rm NLO}_{\Upsilon(11020)\to \Upsilon (2S)}=1.1(\pm 0.5)_{\alpha_s}~{\rm MeV}\,.\label{s:siw:e4}
\end{align}
Finally it is interesting to notice that the sum of semi-inclusive widths for $\Gamma^{\rm LO+NLO}_{\Upsilon(11020)}=28.4\pm 9.4$~MeV is compatible with the experimental value of the total width $\Gamma^{\rm exp}_{\Upsilon(11020)}=24^{+8}_{-6}$~MeV. This is a strong indication that $\Upsilon(11020)$ is a hybrid bottomonium state.

\section{Conclusions} \label{s4}

We reported on a computation, presented in Ref.~\cite{TarrusCastella:2021pld}, in which we computed the transitions of $\Upsilon(10753)$ and $\Upsilon(11020)$ into standard quarkonium and some light quark hadrons under the assumption that they are bottomonium hybrid.

\section*{Acknowledgments}                             

J.T.C. acknowledges financial support by National Science Foundation (PHY-2013184). 

\bibliography{hybtrans}

\end{document}